\documentclass[%
reprint,
 amsmath,amssymb,
 prl,
floatfix,
]{revtex4-1}

\usepackage{graphicx}
\usepackage{dcolumn}
\usepackage{bm}
\usepackage{epstopdf}
\usepackage{color}

\def\corr#1{{{#1}}}     

\newcommand{\la}{\left<}
\newcommand{\ra}{\right>}

\newcommand{\bnperp}{\boldsymbol{\nabla}_\perp}
\newcommand{\Pm}{\text{Pm}}
\newcommand{\Rm}{\text{Rm}}

\begin{document}

\preprint{APS/123-QED}

\title{Transition to turbulent dynamo saturation}

\author{Kannabiran Seshasayanan${}^1$}
 \email{skannabiran@lps.ens.fr}
\author{Basile Gallet${}^2$}
\author{Alexandros Alexakis${}^1$}%
\affiliation{%
 ${}^1$ Laboratoire de Physique Statistique, {\'E}cole Normale Sup{\'e}rieure, CNRS UMR 8550, Universit{\'e} Paris Diderot, Universit{\'e} Pierre et Marie Curie, 24 rue Lhomond, 75005 Paris, France,}
\affiliation{ ${}^2$ Service de Physique de l'\'Etat Condens\'e, CEA,
CNRS UMR 3680, Universit\'e Paris-Saclay, CEA Saclay, 91191 Gif-sur-Yvette,
France.}%

\date{\today}

\begin{abstract}
While the saturated magnetic energy is independent of viscosity in dynamo experiments, it remains viscosity-dependent in state-of-the-art 3D direct numerical simulations (DNS). Extrapolating such viscous scaling-laws to realistic parameter values leads to an underestimation of the magnetic energy by several orders of magnitude. The origin of this discrepancy is that fully 3D DNS cannot reach low enough values of the magnetic Prandtl number $\Pm$. To bypass this limitation and investigate dynamo saturation at very low $\Pm$, we focus on the vicinity of the dynamo threshold in a rapidly rotating flow: the velocity field then depends on two spatial coordinates only, while the magnetic field consists of a single Fourier mode in the third direction. We perform numerical simulations of the resulting set of reduced equations for $\Pm$ \corr{down to $2\cdot10^{-5}$}. This parameter regime is currently out of reach to fully 3D DNS. We show that the magnetic energy transitions from a high-$\Pm$ viscous scaling regime to a low-$\Pm$ turbulent scaling regime, the latter being independent of viscosity. The transition to the turbulent saturation regime occurs at a low value of the magnetic Prandtl number, $\Pm \simeq 10^{-3}$, which explains why it has been overlooked by numerical studies so far.
\end{abstract}

\pacs{Valid PACS appear here}
\maketitle



The magnetic field of most astrophysical objects is believed to originate from the dynamo effect, an instability that converts part of the fluid kinetic energy into magnetic energy. 
The dynamo instability sets in when the flow is sufficiently vigorous to amplify magnetic field perturbations through electromagnetic induction and overcome Ohmic diffusion. In dimensionless form, this happens above a critical value $\Rm_c$ of the magnetic Reynolds number $\Rm=U\ell/\eta$, where $U$ and $\ell$ are the typical velocity and length scales of the flow and $\eta = 1/ \mu_0 \sigma$ is the magnetic diffusivity, with $\sigma$ the electrical conductivity of the fluid and $\mu_0$ the magnetic permeability of vacuum. 
An immediate difficulty arises from the low value of the magnetic Prandtl number $\Pm=\nu / \eta$, where $\nu$ is kinematic viscosity: values of $\Pm$ of the order of $10^{-5}$ are typical of liquid metals and solar system objects. As a consequence, when the flow reaches the $\mathcal{O}(1)$ threshold value $\Rm_c$, the kinetic Reynolds number $\text{Re}=U\ell / \nu$ is in the range $10^5-10^6$ and the flow is fully turbulent. This constitutes a challenge both experimentally and numerically: because of the high power needed to sustain a turbulent flow above $\Rm_c$, large experimental facilities are needed, and only three such experiments have succeeded in producing dynamo magnetic fields \cite{stieglitz2001experimental, gailitis2001magnetic, monchaux2007generation}. On the numerical side, direct numerical simulations (DNS) of the dynamo effect at realistic $\Pm$ values would require gigantic computational resources to accurately resolve the small scales of the fully turbulent flow. State-of-the-art dynamo DNS
are therefore restricted to moderately low values of $\Pm$, typically $\Pm \ge 0.01$ in triple periodic boxes \cite{mininni2007inverse} and $\Pm \ge 0.05$ in spherical geometry \cite{christensen2006scaling,christensen2010dynamo,jones2011planetary,king2013flow}.

Deriving and testing scaling laws that extrapolate numerical results to the physically motivated values of $\Pm$ is essential for relating DNS with observations 
\cite{christensen2006scaling,davidson2013scaling,king2013flow,oruba2014predictive}.
In this quest for scaling relations, probing the dynamo effect at much lower $\Pm$ values is highly desirable, because the turbulence of the background flow strongly affects the magnetic energy produced above threshold. Consider for instance the vicinity of a supercritical dynamo bifurcation, a regime which is relevant to all dynamo laboratory experiments and possibly some planets: close to onset, one expects the magnetic energy to scale linearly with the departure from threshold ($\Rm - \Rm_c$), with  a dimensional prefactor that crucially depends on the value of the magnetic Prandtl number. Indeed, high-$\Pm$ dynamos and theoretical examples of laminar dynamo flows saturate through a balance between the Lorentz force and the viscous one. The magnetic energy above threshold then follows the ``viscous'' scaling-law \cite{petrelis2001saturation}:
\begin{equation}
\frac{|{\bf B}|^2 \ell^2}{ \rho \mu_0 \eta^2} \propto \Pm \, (\Rm - \Rm_c)  \, . \label{viscousscaling}
\end{equation}
By contrast, laboratory experiments indicate that this saturation is achieved through a balance between the Lorentz force and the nonlinear advective term of the Navier-Stokes equation \cite{petrelis2007GAFD}. This leads to the ``turbulent'' scaling-law for the magnetic energy \cite{petrelis2001saturation}:
\begin{equation}
\frac{|{\bf B}|^2 \ell^2}{ \rho \mu_0 \eta^2} \propto (\Rm - \Rm_c)  \, , \label{turbscaling}
\end{equation}
which corresponds to a much higher magnetic energy than the viscous scaling law (\ref{viscousscaling}) by a factor $\Pm^{-1}$.

\corr{An interesting approach to test these theoretical predictions is the one based on shell models of MHD turbulence. Such phenomenological models are not meant to quantitatively describe the flows, but rather to capture their statistical properties at reasonable computational cost \cite{Plunian2013}: they provide evidence for the scaling-law (\ref{turbscaling}) when the magnetic Prandtl number is low enough, $\Pm \lesssim 1$ \cite{Stepanov2006}.}

\corr{The natural next step towards quantitative numerical dynamo models would be to reproduce the turbulent scaling regime directly from numerical solutions of the Navier-Stokes and induction equations, a task which remains beyond reach of state-of-the-art fully 3D DNS. Indeed, it} has been recently shown that all spherical dynamo simulations obey the viscous scaling law \cite{king2013flow, oruba2014predictive}: because of the moderately low $\Pm$, they are restricted to the  viscosity-dominated regime, which makes their extrapolation to Earth like parameters questionable. A central question is therefore: how much smaller need the magnetic Prandtl number be to start observing the turbulent scaling regime (\ref{turbscaling})? 

We address this question by focusing on rapidly rotating flows in the vicinity of the dynamo threshold. 
Global rotation is a relevant ingredient of many planetary dynamos, which strongly affects the dynamo characteristics \cite{donati2004rotation}. 
For rapid global rotation, we are able to reduce the full MHD system to a set of quasi-2D equations governing the interaction between a two-dimensional three-component (2D3C) flow and a vertically-dependent dynamo magnetic field. This approach allows us to bypass the current limitations of 3D DNS: we focus on the high-Reynolds-number regime where nonlinear advection strongly dominates over viscous effects, in a quasi-2D system of manageable computational cost.

\begin{figure}[!htb]
\includegraphics[width=0.5\textwidth]{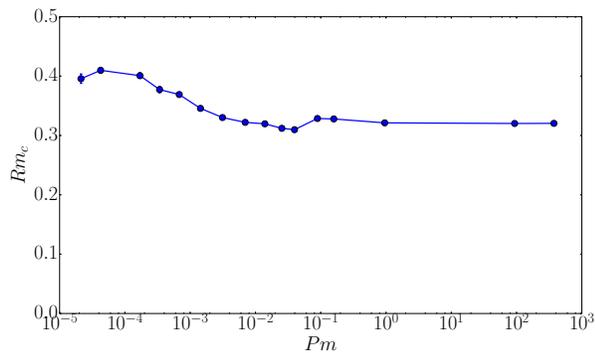}
\caption{Threshold magnetic Reynolds number $\Rm_c$ for dynamo action as a function of $\Pm$, for $\gamma \ell^2 / \eta = 5.1 \, 10^{-3}$.
\label{fig:RmcPm} }
\end{figure}

{\it Reduced equations.} We consider a flow driven by a vertically invariant body force ${\bf f}(x,y)$ in a frame rotating at a rate $\Omega$ around the vertical $z$ axis. It was recently proven that the corresponding 2D turbulent flow is stable to 3D perturbations provided the Rossby number is sufficiently low \cite{gallet2015exact}. We focus on this parameter range, the turbulent 2D3C velocity field ${\bf u}=[u(x,y,t), v(x,y,t), w(x,y,t)]$ being the base flow of the present dynamo study. For the kinematic dynamo problem, the invariance of the flow along the vertical direction allows us to decompose the 
magnetic field into vertical Fourier modes and to study their evolution independently:
\begin{equation}
{\bf B}(x,y,z,t)={\bf b}(x,y,t) e^{i k z} + c.c. \, , \label{Fourier}
\end{equation}
where $c.c.$ denotes the complex conjugate and $k$ the vertical wavenumber. Inserting such a Fourier mode into the induction equation leads to the evolution equation for ${\bf b}$:
\begin{equation}
\partial_t {\bf b}= (\bnperp + i k {\bf e}_z) \times ({\bf u} \times {\bf b}) + \eta (\bnperp^2 - k^2) {\bf b}\, , \label{induction}
\end{equation}
where $\bnperp=(\partial_x, \partial_y, 0)$ and ${\bf e}_z$ is the unit vector along $z$ \cite{smith2004vortex,seshasayanan2015turbulent,seshasayanan2016Kazantsev}.

We focus on the weakly nonlinear regime in the vicinity of the dynamo threshold, for which we can keep only the first unstable magnetic mode of the form (\ref{Fourier}). The instability saturates through the action of the Lorentz force. The latter being quadratic in the magnetic field (\ref{Fourier}), it contains two harmonics in $z$:
\begin{itemize}
\item A $z$-independent component (harmonic $0$), through which the magnetic field directly feeds back onto the $z$-invariant base flow.
\item A second harmonic component, with vertical dependence $e^{\pm 2 i k z}$. It forces a second harmonic in the velocity field, ${\bf u}_2 = \hat{{\bf u}}_2(x,y,t) e^{2i k z} + c.c.$, the amplitude of which results from a balance between the Coriolis and Lorentz forces:  
\begin{equation}
{\bf u}_2 \sim \frac{{\bf B}^2}{\rho \mu_0 \ell \Omega} \, . \label{equ2}
\end{equation}
\end{itemize}
The vertical average of the Coriolis force being absorbed by the pressure gradient, the $z$-independent flow follows the equation:
\begin{eqnarray}
\partial_t {\bf u} +  ({\bf u} \cdot \bnperp) {\bf u} & = & -\bnperp p - \gamma {\bf u}_\perp + \nu \bnperp^2 {\bf u} + {\bf f}(x,y) \label{NS2D} \\
\nonumber & + & \frac{1}{\rho \mu_0} \left[  \left[ (\bnperp + i k {\bf e}_z) \times {\bf b} \right] \times {\bf b}^*  + c.c. \right] \, , 
\end{eqnarray}
where ${\bf u}_\perp=(u,v,0)$, $\gamma$ is a linear Ekman friction coefficient \corr{\cite{Pedloskybook}}, and the last term is the vertical average of the Lorentz force. The latter induces a correction to the turbulent 2D3C base flow of order ${\bf B}^2 / (\rho \mu_0 U)$. By contrast, equation (\ref{equ2}) indicates that ${\bf u}_2$ is smaller than this correction by a factor equal to the Rossby number $U / {\ell \Omega}$, which is asymptotically small in the present study: we therefore discard ${\bf u}_2$ in the following. The weakly nonlinear regime in the vicinity of the dynamo threshold then corresponds to the interaction between the first unstable vertical Fourier mode of the magnetic field and the $z$-invariant 2D3C flow. Their coupled evolution obeys the reduced system of equations (\ref{induction}-\ref{NS2D}).

{\it Numerical experiments.} We solve equations (\ref{induction}-\ref{NS2D}) inside a domain $(x,y) \in [0, 2 \pi L]^2$ using standard pseudo-spectral methods \cite{Gomez05, seshasayanan2015turbulent}. The body-force has the Roberts-flow geometry:
\begin{equation}
{\bf f}(x,y)=F \, [\cos(y/\ell), \, \sin(x/\ell), \, \cos(x/\ell)+\sin(y/\ell)] \, ,
\end{equation}
where the scale of the forcing is $\ell=L/4$. \corr{This value was shown to be close to optimal for reducing the dynamo threshold \cite{Sadek2016Optimal}, which is desirable to reach the low-magnetic-Prandtl-number regime.}

\corr{The vertical wavenumber is set to $k=0.2 /L$, which corresponds to the lowest wavenumber inside a domain of height $10 \pi \, L$. The influence of this parameter on the dynamo threshold and magnetic field geometry was studied in detail in a previous publication \cite{seshasayanan2015turbulent}, the phenomenology being that of standard $\alpha^2$ dynamos \cite{Moffattbook,Ponty2011}.}


\begin{figure}[]
                \includegraphics[scale=0.22 ]{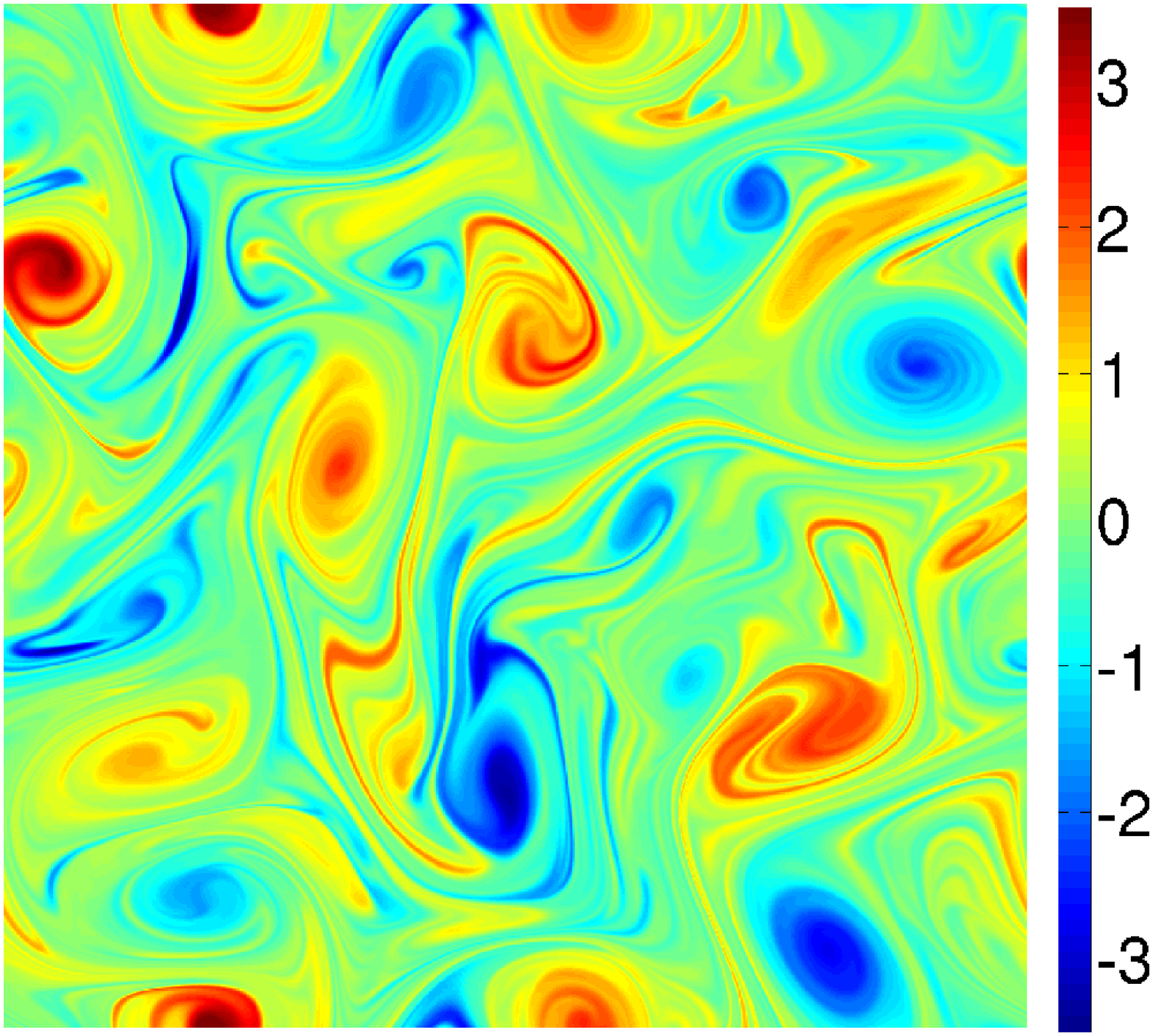} \\
\hspace{0.0 cm} \includegraphics[scale=0.22 ]{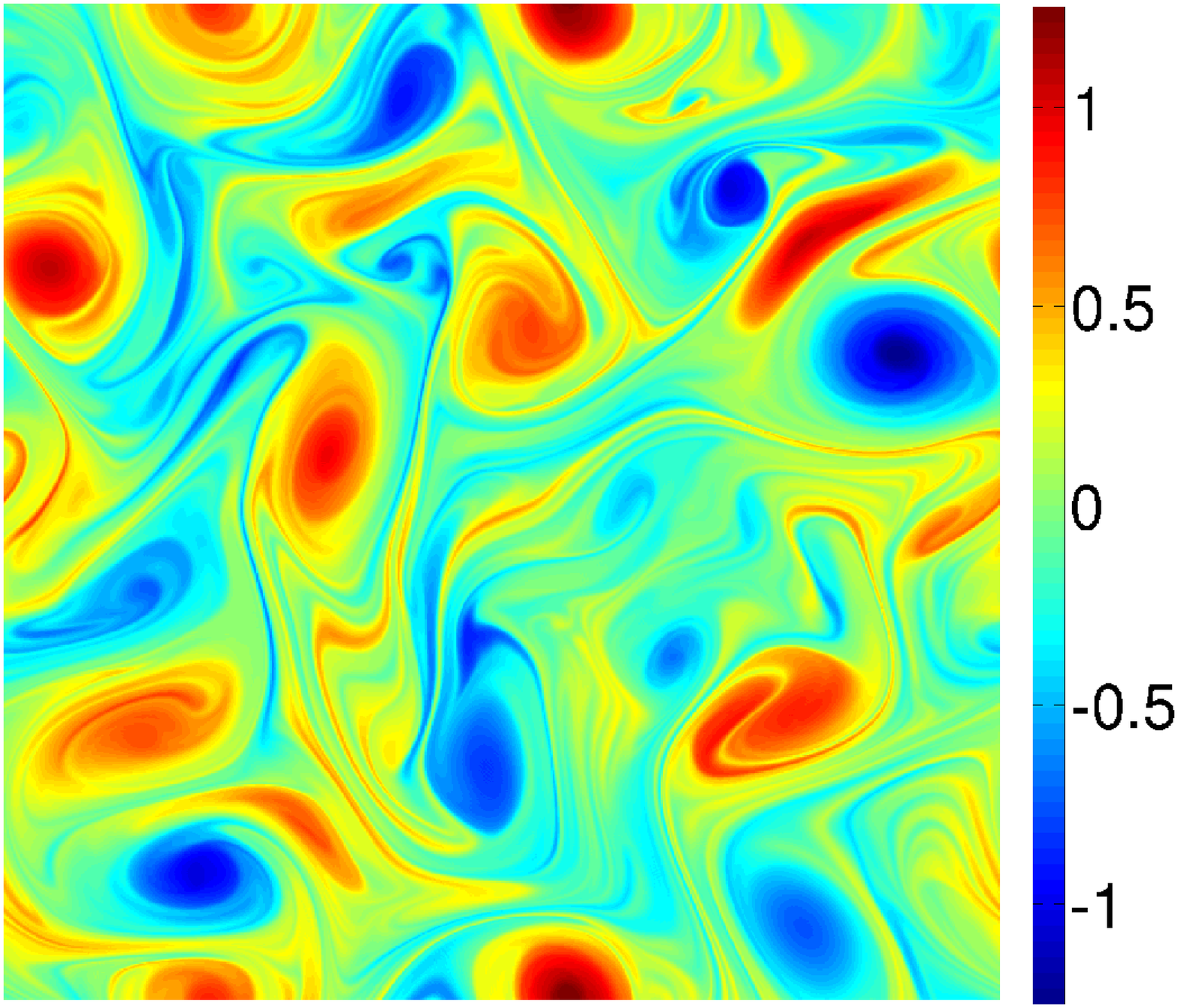} \\
\hspace{0.0 cm} \includegraphics[scale=0.22 ]{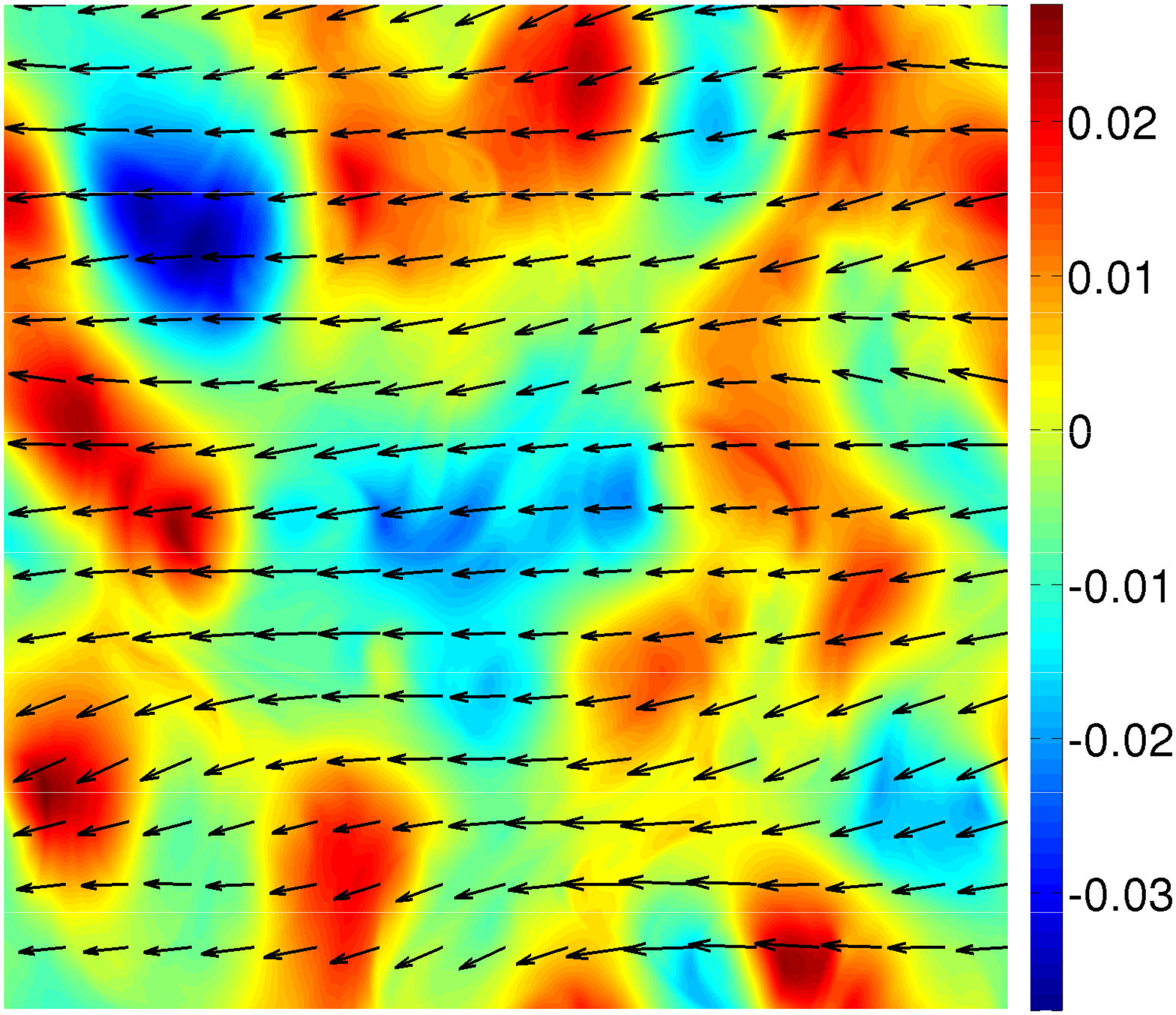}
\caption{Snapshots of the saturated state for $\Pm=4.25 \cdot 10^{-5}$, $\gamma \ell^2 / \eta = 5.1 \cdot 10^{-3}$ and $Rm=0.44$. {\bf Top:} vertical vorticity in units of $U/\ell$. {\bf Middle:} vertical velocity in units of $U$. {\bf Bottom:} \corr{dimensionless magnetic field  ${\bf B} \, \ell / \sqrt{\rho \mu_0 \eta^2}$ in the plane $z=0$. The arrows indicate the horizontal components while color codes for the vertical one. These arrows correspond to a typical magnitude $4 \cdot 10^{-2}$ of the dimensionless horizontal magnetic field. Their average direction rotates with $z$.} \label{fig:snapshots} }
\end{figure}

We first solve the purely hydrodynamic problem until a statistically steady state is reached.
We denote as $U$ the root-mean-square velocity of this state that depends on $F,\nu,\gamma$.
This flow is then used as the starting point of the MHD simulations.
We solve the MHD problem for increasing values of $F$, i.e., for increasing values of the magnetic Reynolds number $\Rm$
while keeping $\Pm$ fixed ($\nu, \gamma$ and $\eta$ are fixed). Above the threshold value $\Rm_c$ for dynamo action, the dynamo instability sets in and magnetic perturbations grow. 
As shown in figure \ref{fig:RmcPm}, $\Rm_c$ varies little with $\Pm$, 
with $\Rm_c \in [0.30 ; 0.42]$ over seven decades of $\Pm$. \corr{Such a constant value of $\Rm_c$ at low $\Pm$ is rather generic and has been reported for several other forcing geometries \cite{Ponty2005,Iskakov2007}. By contrast, the behavior of $\Rm_c$ in the transitional region of intermediate $\Pm$ strongly depends on the structure of the forcing: for some fully 3D flows with weak scale separation, $\Rm_c$ displays a two-fold increase at intermediate $\Pm$ \cite{Ponty2011}. 
The weak variation of $\Rm_c$ with $\Pm$ in figure \ref{fig:RmcPm} is therefore attributed to both the scale separation and the 2D3C nature of the flow \cite{Sadek2016Optimal,seshasayanan2017onset}.}

\begin{figure}[]
\includegraphics[scale=0.42]{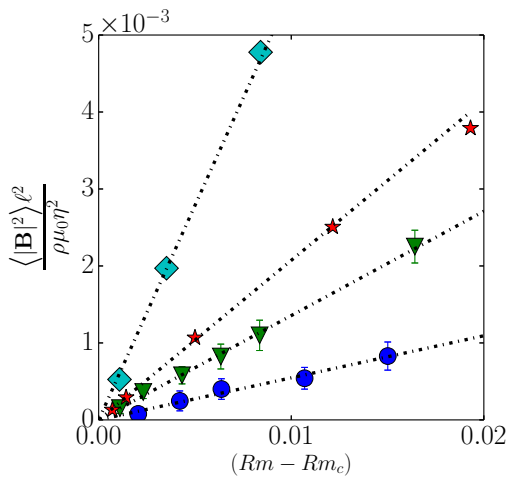}
\caption{\label{fig:bifurcation} Magnetic energy as a function of the departure from onset, for several values of the magnetic Prandtl number. 
The friction is $\gamma \ell^2 / \eta = 5.1 \, 10^{-3}$ and symbols are: $\bullet$, $\Pm=1.4 \, 10^{-3}$; $\triangledown$, $\Pm=7.0 \, 10^{-3}$; $\star$, $\Pm=4.0 \, 10^{-2}$; $\diamond$, $\Pm=8.9 \, 10^{-2}$. The dashed linear fits allow to extract the slope $S(\Pm)$ of each bifurcation curve.}
\end{figure}

After the initial growth phase, the magnetic field saturates through the feedback of the Lorentz force onto the 2D3C flow. In figure \ref{fig:snapshots}, we show snapshots of the velocity and magnetic fields in the saturated state, for $\Pm=4.25 \cdot 10^{-5}$. In agreement with standard $\alpha^2$-dynamo theory \cite{Moffattbook}, the vertical velocity and vorticity are concentrated at the forcing scale $\ell$ \corr{while the dynamo magnetic field is at large scale, with horizontal components more energetic than the vertical one}. We compute the space and time averaged magnetic energy in the saturated state, to produce bifurcation curves such as the ones shown in figure \ref{fig:bifurcation}. We repeat this procedure for various values of the magnetic Prandtl number $\Pm$, and from each bifurcation curve we extract the slope $S(\Pm)$ relating the magnetic energy to the departure from onset:
\begin{equation}
\frac{\la |{\bf B}|^2 \ra \ell^2}{\rho \mu_0 \eta^2} = S(\Pm) \times  (\Rm-\Rm_c) \, ,
\end{equation}
where the angular brackets denote 3D space and time average. $S(\Pm)$ is the central quantity of the present study.

\begin{figure}[!htb]
\includegraphics[scale=0.43]{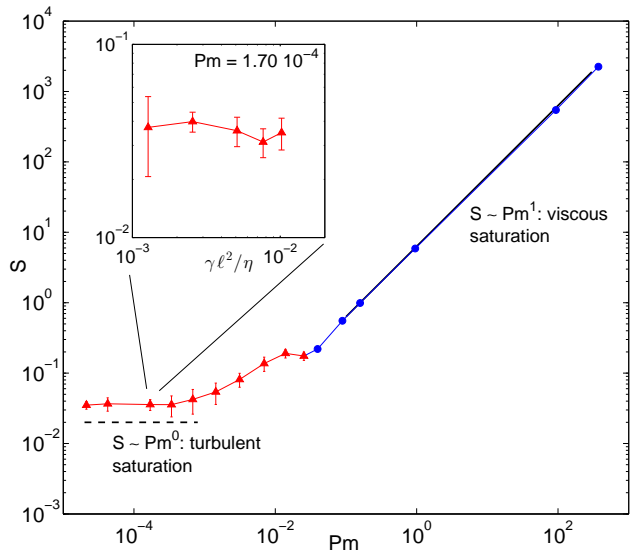}
\caption{\label{fig:slopes} Slope $S$ of the magnetic energy above onset. High-$\Pm$ solutions are time-independent (blue circles) and obey the quantitative prediction (\ref{viscousquench}) from viscous alpha quenching (thick solid line). For low $Pm$, the flow is time-dependent (red triangles). For $Pm \lesssim 10^{-3}$, the dynamo saturation follows the turbulent scaling-law (\ref{turbscaling}), represented as a dashed eye-guide. The main figure corresponds to a friction coefficient $\gamma \ell^2/\eta=5.1 \, 10^{-3}$. The inset highlights the independence of $S$ on friction in the turbulent saturation regime. }
\end{figure}

{\it From viscous to turbulent saturation.} In figure \ref{fig:slopes}, we show $S(\Pm)$ over seven decades of $\Pm$. For large $\Pm$, the flow is laminar and has a low Reynolds number near the dynamo threshold. Accordingly, the dynamo saturation obeys the viscous scaling-law (\ref{viscousscaling}), i.e., $S (\Pm) \sim \Pm$ for large $\Pm$. The precise expression of $S(\Pm)$ can be determined analytically in the limit of scale separation $k \ell \ll 1$ and corresponds to the usual viscous alpha quenching. For small friction $\gamma \ell^2 \ll \nu$, following the standard weakly nonlinear approach \cite{gilbert1990inverse, fauve2003} we obtain:
\begin{equation}
S(\Pm)= \sqrt{\frac{2}{k \ell}} \, \Pm  \, . \label{viscousquench}
\end{equation}

%
%
This analytical prediction is displayed in figure \ref{fig:slopes} with a solid line and captures perfectly the high-$\Pm$ behavior of $S(\Pm)$.

In contrast with such viscous dynamos, $S(\Pm)$ reaches a plateau at low $\Pm$, with values orders of magnitude larger than predicted by the laminar theory. This corresponds to the turbulent scaling-law (\ref{turbscaling}), for which $S(\Pm)$ is independent of $\Pm$. More precisely, in this regime $S$ is independent of both $\nu$ and $\gamma$ (see inset of figure \ref{fig:slopes}) and the dominant balance in the Navier-Stokes equation (\ref{NS2D}) is between the Lorentz force and the nonlinear term. To our knowledge, this study constitutes the first numerical observation of the turbulent scaling regime of dynamo saturation \corr{in a solution of the coupled Navier-Stokes and induction equations}. This is because extremely low values of $\Pm$ are needed to observe such turbulent saturation: the plateau in figure \ref{fig:slopes} arises for $\Pm \lesssim 10^{-3}$, an order of magnitude below the smallest values of $\Pm$ achieved in state-of-the-art fully 3D DNS.

%

{\it Discussion.} 
\corr{We have reported the numerical observation of the turbulent scaling regime for dynamo saturation in a solution of the MHD equations. Our work therefore quantifies for the first time previous theoretical estimates based on dimensional analysis \cite{petrelis2001saturation} and shell models \cite{Stepanov2006}:}
in the present setup, the turbulent scaling regime sets in for values of the magnetic Prandtl number an order of magnitude lower than currently achieved by state-of-the-art fully 3D DNS. 
This explains the mismatch between spherical dynamo simulations, which obey the viscous scaling law \cite{oruba2014predictive}, and experimental dynamos, which follow the turbulent one \cite{petrelis2007GAFD}. 
\corr{Because the turbulent scaling regime is likely to be the generic situation for many natural and experimental dynamos, this study  paves} the way for quantitative numerical estimates of the magnetic field in astrophysical objects and laboratory experiments. 

\begin{acknowledgments}
This work was granted access to the HPC resources of GENCI-CINES (Project
No.x2014056421, x2015056421)
and MesoPSL financed by the Region Ile de France and the project Equip\@Meso
ANR-10-EQPX-29-01. KS acknowledges support from LabEx ENS-ICFP:
ANR-10-LABX-0010/ANR-10-IDEX-0001-02 PSL and from {\'E}cole Doctorale {\^I}le de France (EDPIF). BG acknowledges support from LabEx PALM ANR-10-LABX-0039.
\end{acknowledgments}


\bibliography{apssamp}

\end{document}